\documentclass[preprint,3p,times,fleqn]{elsarticle}

\usepackage{amsmath}
\usepackage{amssymb}
\usepackage{psfrag}
\usepackage{color}
\usepackage{stfloats}
\usepackage{mathrsfs}
\usepackage{tabularx}
\usepackage{booktabs}
\usepackage{colortbl}
\usepackage{rotating}
\usepackage{boldline}
\usepackage{bm}
\usepackage{changes}
\usepackage{url}

\usepackage{siunitx}

\usepackage{float}
\usepackage{lineno}

\usepackage{multirow}

\usepackage{hyperref}

\journal{Engineering with Computers}
\begin{document}

\renewcommand{\topfraction}{0.98}	
\renewcommand{\bottomfraction}{0.98}
\setcounter{topnumber}{3}
\setcounter{bottomnumber}{3}
\setcounter{totalnumber}{4}         
\setcounter{dbltopnumber}{4}        
\renewcommand{\dbltopfraction}{0.98}	
\renewcommand{\textfraction}{0.05}	
\renewcommand{\floatpagefraction}{0.5}	
\renewcommand{\dblfloatpagefraction}{0.5}	
\newcommand{\beq}{\begin{equation}}
\newcommand{\eeq}{\end{equation}}
\newcommand{\divg}{\mbox{\rm{div}}\,}
\newcommand{\Divg}{\mbox{\rm{Div}}\,}
\newcommand{\D}  {\displaystyle}
\newcommand{\DS} {\displaystyle}
\newcommand{\RM}[1]{\textit{\MakeUppercase{\romannumeral #1{}}}}
\newtheorem{remark}{\bf{{Remark}}}
\def\sca   #1{\mbox{\rm{#1}}{}}
\def\mat   #1{\mbox{\bf #1}{}}
\def\vec   #1{\mbox{\boldmath $#1$}{}}
\def\scas  #1{\mbox{{\scriptsize{${\rm{#1}}$}}}{}}
\def\scaf  #1{\mbox{{\tiny{${\rm{#1}}$}}}{}}
\def\vecs  #1{\mbox{\boldmath{\scriptsize{$#1$}}}{}}
\def\tens  #1{\mbox{\boldmath{\scriptsize{$#1$}}}{}}
\def\tenf  #1{\mbox{{\sffamily{\bfseries {#1}}}}}
\def\ten   #1{\mbox{\boldmath $#1$}{}}
\def\Ass  {\overset{\hspace*{0.4cm} n_{\scas{el}}}
          {\underset{\scaf{c},\scaf{d}=1}{\msf{A}}}}
\def\ltr   #1{\mbox{\sf{#1}}}
\def\bltr  #1{\mbox{\sffamily{\bfseries{{#1}}}}}

\sloppy
\begin{frontmatter}
\title{\Large WarpPINN-fibers: improved cardiac strain estimation from cine-MR with physics-informed neural networks}

\author[01]{Felipe~\'Alvarez Barrientos}
\author[02]{Tom\'as~Banduc}
\author[01]{Isabeau Sirven}
\author[02,03,04]{Francisco~Sahli Costabal\corref{cor1}}
\ead{fsc@ing.puc.cl}

\cortext[cor1]{Corresponding author}

\address[01]{\'Ecole Polytechnique, Palaiseau, France}

\address[02]{Millennium Institute for Intelligent Healthcare Engineering, iHEALTH}

\address[03]{Department of Mechanical and Metallurgical Engineering, School of Engineering, Pontificia Universidad Cat\'olica de Chile, Santiago, Chile}

\address[04]{Institute for Biological and Medical Engineering, Schools of Engineering, Medicine and Biological Sciences, Pontificia Universidad Cat\'olica de Chile, Santiago, Chile}

\begin{abstract} %
The contractile motion of the heart is strongly determined by the distribution of the fibers that constitute cardiac tissue. Strain analysis informed with the orientation of fibers allows to describe several pathologies that are typically associated with impaired mechanics of the myocardium, such as cardiovascular disease. Several methods have been developed to estimate strain-derived metrics from traditional imaging techniques. However, the physical models underlying these methods do not include fiber mechanics, restricting their capacity to accurately explain cardiac function. In this work, we introduce WarpPINN-fibers, a physics-informed neural network framework to accurately obtain cardiac motion and strains enhanced by fiber information. We train our neural network to satisfy a hyper-elastic model and promote fiber contraction with the goal to predict the deformation field of the heart from cine magnetic resonance images. For this purpose, we build a loss function composed of three terms: a data-similarity loss between the reference and the warped template images, a regularizer enforcing near-incompressibility of cardiac tissue and a fiber-stretch penalization that controls strain in the direction of synthetically produced fibers. We show that our neural network improves the former WarpPINN model and effectively controls fiber stretch in a synthetic phantom experiment. Then, we demonstrate that WarpPINN-fibers outperforms alternative methodologies in landmark-tracking and strain curve prediction for a cine-MRI benchmark with a cohort of 15 healthy volunteers. 
We expect that our method will enable a more precise quantification of cardiac strains through accurate deformation fields that are consistent with fiber physiology, without requiring imaging techniques more sophisticated than MRI. 
\end{abstract}

\end{frontmatter}


\section{Introduction}\label{sec1}

Cardiovascular disease (CVD) is a leading cause of death worldwide, accounting for 32\% of all deaths globally \cite{who2025cvd}. CVDs are prevalent within the world population at epidemic scale and span a variety of disorders of the circulatory system characterized by alterations in the rhythm of the heart, disrupted hemodynamics and the development of mechanistic abnormalities in the cardiac muscle and the blood vessels irrigating the body. The increased lifetime risk of CVD is associated with the global burden of metabolic syndrome \cite{ZHANG2024155999,wang2023metabolic,seth2025}. The rising prevalence of CVDs in developing countries and the deterioration in the quality of life of affected patients motivate public health initiatives and research for the accurate assessment of cardiac function, the refinement of diagnostic tools and the development of cost-effective technologies for their treatment \cite{gersh2010,soloveva2023hrqol,hcaf022,mensah2023,seth2025}.\newline 

CVDs typically emerge from structural changes of the cardiac muscle, which translate into the motion impairment of the heart. The mechanistic nature of cardiac dysfunction allows us to determine the initiation and progression of several CVDs by the use of pressure markers and measurements of tensile and compressive stress of tissue \cite{seth2025, SUCOSKY202023}, such is the case for cardiomyopathy \cite{Lin2016}, pulmonary hypertension \cite{vonSiebenthal2016} and myocardial infarction \cite{Korosoglou2024, Voorhees2014}. Overall, strain-based metrics unveil subclinical conditions that tend to be undetectable by standard quantities, thus allowing to differentiate between multiple pathologies \cite{Steen2021, ZHANG2021104431}.  


The contractility of the heart is determined by the orientation of cardiac fibers \cite{LI2025107025}. For the case of the ventricles, fibers are arranged in such a way that strain along their direction is homogenized \cite{Rijcken1997} and ejection fraction is locally maximized \cite{Sallin1969}. This makes cardiac fibers circularly oriented within the myocardium and helically distributed near the endocardial and epicardial layers of the ventricles \cite{Sallin1969}. 

The anisotropic ensemble of cardiac muscle allows to describe cardiac motion in patients with CVD \cite{Voorhees2014,Zhang2015,RodriguezPadilla2022}. For this reason, strain metrics that encode the structural constraints imposed by myofiber orientation provide meaningful information to improve risk stratification and to accurately explain cardiac function at a multilevel scale \cite{Moulin2021}. 



To recover cardiac motion non-invasively, magnetic resonance imaging (MRI) techniques enhanced with registration algorithms have enabled the reconstruction of displacement fields without increasing scan times \cite{lopez2023warppinn,de2012temporal,morales2021deepstrain,mansi2011ilogdemons,mauger2018}. 
In this context, deep learning approaches have shown to be effective in the task of landmark tracking and also consistent with the observed dynamics across the cardiac cycle \cite{de2012temporal,mansi2011ilogdemons}. However, these methods do not usually account for the mechanical processes involved in cardiac motion, and fail to properly estimate physiological strains of tissue \cite{lopez2023warppinn}. To incorporate the mechanical properties of the myocardium into the image registration task, the WarpPINN model has been introduced as a physics-informed neural network approach that enforces quasi-incompressibility to a hyper-elastic model of cardiac tissue \cite{lopez2023warppinn}. This method accurately predicts landmark deformation, achieves competitive performance to both numerical methods and deep learning strategies for the reconstruction of displacement fields and generates smooth, non-drifting curves for strain, with more realistic radial values \cite{lopez2023warppinn}. However, the WarpPINN method still requires refinement to fully capture the true mechanics involved in cardiac motion, since it is biased towards isotropic deformation, as the hyper-elastic regularization does not account for the local action of myofibers during contraction.


Diffusion tensor imaging (DTI) techniques have been developed to non-invasively reconstruct fiber orientation in patient-specific anatomies. Changes in the transmural variation of DTI fibers have been linked to several diseases, such as dilated cardiomyopathy \cite{vondeuster2016}, hypertrophic cardiomyopathy \cite{khalique2020}, cardiac amyloidosis \cite{khalique2020,gotschy2019} and aortic valve stenosis \cite{gotschy2021}. Moreover, the reorientation of cardiomyocyte aggregate revealed by DTI serves as a marker for the onset and progression of myocardial remodeling in aortic valve stenosis \cite{gotschy2021}.

DTI methods pose technical challenges in terms of acquisition times and artifacting due to cardiac and respiratory motion for \textit{in vivo} measurements \cite{Scollan1998, Scollan2000, zhukov}. Moreover, DTI sequences are commonly restricted to a reduced amount of slices---usually three---and are not routinely acquired in clinical practice, making the complete reconstruction of myocardial fiber orientation unattainable in real-life scenarios. Automated algorithms supplied with electro-anatomical data \cite{RONEY2019278}, physics-informed deep learning architectures \cite{RuizHerrera2022} and statistically-driven reconstruction methodologies \cite{LEKADIR2016105} have been developed to tackle these limitations, but they still require additional information that is not accessible through standard MR data.

In this work, we extend the WarpPINN workflow to an enhanced mechanical model by introducing WarpPINN-fibers, a fully connected neural network architecture that incorporates the hyper-elastic model to the physics of the myocardium while favoring diffeomorphic deformation maps and imposing physiological constraints defined by the mechanical behavior of myofibers. The proposed method aims to minimize strain in the direction of computationally generated fibers, with the objective to achieve more realistic deformations of the cardiac muscle in the image registration problem for cine-MRI without any additional measurements.

The manuscript is structured as follows: in section 2 we introduce the methodology to solve the image registration problem. Here, we present a general framework to recover deformation fields from MR images, review the WarpPINN architecture and propose the WarpPINN-fibers model. We also describe a rule-based method to generate fibers, the Fourier feature mapping and elaborate on the workflow to train our neural network. Finally, we present the experimental setting used to validate the WarpPINN-fibers approach, where we include a synthetic example and an applied case with real patients. Section 3 is devoted to showing the results of the synthetic example and the applied case, where we compare the predictions of our neural network to WarpPINN. In this section we additionally compare WarpPINN-fibers to alternative methods for landmark tracking and strain curve generation. Finally, in section 4, we discuss the results, highlight the limitations of our study and conclude on the effectiveness of our method. 

\section{Methods}\label{sec2}

\subsection{Registration of cine MRI}
To estimate cardiac strain from cine MR images, we define a image registration problem to a sequence of frames obtained with SSFP MRI. Here, we aim to find a set of transformations such that a sequence of images acquired during the cardiac cycle can be mapped pixel by pixel to a reference state at all times. 

First, we consider two images displaying the same object: a reference image $R:\Omega_0\to\mathbb{R}$ with spatial domain $\Omega_0\subseteq \mathbb{R}^n$ (n=2,3) and a template image $T:\Omega_1\to\mathbb{R}$ with spatial domain $\Omega_1\subseteq \mathbb{R}^n$. We assume that there exists a diffeomorphic map ($f$ is said to be diffeomorphic if it is continuously differentiable and has a continuously differentiable inverse) $\vec{\varphi}:\Omega_0\to\Omega_1$ such that the reference image matches the warped template via $\vec{\varphi}$. That is, $R(\vec{X})=(T\circ\vec{\varphi})(\vec{X})$. Then, the image registration problem is stated as an optimization problem where the goal is to find a minimizer $\vec{\varphi}$ of the functional
\begin{align*}
    \mathcal{J}_0(\vec{\varphi}):=\left|\left|R(\cdot)-(T\circ\vec{\varphi})(\cdot)\right|\right|_p^p+\mu\mathcal{R}(\vec{\varphi})
\end{align*}

for some suitable $p$-norm $||\cdot||_p$. Here, the first term of $\mathcal{J}_0(\vec{\varphi})$ measures the discrepancy between the reference $R$ and the warped template $T\circ\vec{\varphi}$ in the entire spatial domain, while the second term acts as a regularizer that handles the ill-posed nature of the reconstruction problem \cite{lopez2023warppinn}.

When images are displayed by a sequence of snapshots $\{T_{t}\}_{t\in\mathcal{T}}$ recorded at times $t\in \mathcal{T}$, we require to extend the image registration problem from a spatial setting to a spatio-temporal setting. For this purpose, we consider a family of images $T_t:\Omega_t\to \mathbb{R}$ indexed by a time interval $\mathcal{T}:=[t_\text{ed-1},t_\text{ed-2}]$ spanning the cardiac cycle, from end-diastole to end-diastole. Without loss of generality, we set $t_\text{ed-1}=0$ and $t_\text{ed-2}=1$. To our previous assumption regarding the existence of a diffeomorphic deformation field, we additionally postulate that frames acquired within short time intervals vary smoothly. Formally, we assume there exists a differentiable map $\varphi:t\mapsto\mathcal{C}^1(\Omega_0,\Omega_t)$ such that $\varphi(t):\Omega_{0}\to\Omega_{t}$ is diffeomorphic and the equality $T_{0}=T_t\circ\varphi(t)$ holds point-wise. We denote $\varphi(t)(\vec{X})$ by $\vec{\varphi}(t,\vec{X})$ and fix $R:=T_{0}$ to be the reference image, while regarding the remaining images $\{T_t\}_{t\in(0,1]}$ as templates. Then, we reformulate the image registration task as a spatio-temporal reconstruction problem for the deformation field $\vec{\varphi}$, where the goal is to minimize the functional
\begin{align*}
    \mathcal{J}_1(\vec{\varphi}):= \displaystyle\int\limits_{0}^{1}\left(||R(\cdot)-(T_t\circ\vec{\varphi})(t,\cdot)||^p_p+\mu\mathcal{R}(\vec{\varphi})\right)\text{d}t
\end{align*} 
In the context of cine-MR images, data is constituted by a finite sequence of frames $\{T_{t_i}\}_{i=0}^{N_\text{fr}}\subseteq[0,1]$, with $t_0=0$ and $t_{N_\text{fr}}=1$. Each frame is composed of a discrete stack of 2D images that conform a volumetric representation of the heart. Then, each frame can be represented in 3D space by a structured grid, and the cine images correspond to 3D$+$t data. We consider a mesh with points $\{\vec{X}_j^\text{d}\}_{j=1}^{N_\text{d}}$ for the pixel domain and a uniform grid $t_i := i\Delta t$ for time. Additionally, under the physics-informed framework, we evaluate our regularizer in a set of points $\{(\vec{X}_k^\text{c},t_k^\text{c})\}_{k=1}^{N_\text{c}}$ uniformly sampled from the spatio-temporal domain. Then, our objective is to find a field $\vec{\varphi}$ that minimizes
\begin{align*}
    \displaystyle\sum\limits_{i=1}^{N_\text{fr}}\displaystyle\sum\limits_{j=1}^{N_\text{d}}\left|R(\vec{X}^\text{d}_j)-(T_{t}\circ\vec{\varphi})(t_i,\vec{X}_j^\text{d})\right|^p\Delta\vec{x}\Delta t+\mu\dfrac{1}{N_\text{c}}\displaystyle\sum\limits_{k=1}^{N_\text{c}}F(\varphi(t_k^\text{c},\vec{X}_k^\text{c}))
\end{align*}
Here, we have imposed a regularizer of the form $\mathcal{R}(\vec{\varphi})=\int F(\vec{\varphi})$ and defined $R:=T_{0}$. 

\subsection{Fiber-Informed Neural Network Architecture}
\subsubsection{WarpPINN Model}\label{warpinn}
In our previous work, we aimed to estimate a non-rigid deformation map $\vec{\varphi}(t,\vec{X})$ by the displacement field decomposition $\vec{\varphi}(t,\vec{X})\simeq\vec{X}+\hat{\vec{u}}(t,\vec{X};\theta)=:\vec{\varphi}(t,\vec{X};\theta)$, where $\hat{\vec{u}}(t,\vec{X};\theta)$ is a fully-connected neural network architecture with parameters $\theta$ and a Fourier feature encoding for the spatial coordinates. In this setting, the objective function from the WarpPINN method was comprised of a data similarity loss and a hyper-elastic regularizer enforcing quasi-incompressibility of the myocardium. Namely, for a set of points $\{\vec{X}^\text{d}_j
\}_{j=1}^{N_\text{d}}\subseteq \Omega_{0}$ corresponding to the pixel space of the reference image $R$, a collection of registration times $\{t_i:=i\Delta t\}_{i=1}^{N_\text{fr}}$ and a set of collocation points $\{(t_k^{\text{bg}},\vec{X}_{k}^{\text{bg}})\}_{k=1}^{N_\text{bg}}\cup \{(t_k^{\text{myo}},\vec{X}_{k}^{\text{myo}})\}_{k=1}^{N_{\text{myo}}}\subseteq [0,1]\times\Omega_{0}$ the WarpPINN loss function is defined as

{\small\begin{align*}
    \mathcal{L}_{0}(\theta):=&\dfrac{1}{N_\text{fr}}\displaystyle\sum\limits_{i=1}^{N_\text{fr}}\dfrac{1}{N_\text{d}}\sum\limits_{j=1}^{N_\text{d}}|R(\vec{X}_j^\text{d})-T_{t_i}(\vec{\varphi}(t_i,\vec{X}_j^\text{d};\theta))|\\
    &+\mu\left(\dfrac{1}{N_\text{myo}}\displaystyle\sum\limits_{k=1}^{N_\text{myo}}W\left(\vec{\varphi}\left(t_k^{\text{myo}},\vec{X}_k^{\text{myo}};\theta\right);\lambda_{\text{myo}}\right)+\dfrac{1}{N_\text{bg}}\displaystyle\sum\limits_{k=1}^{N_\text{bg}}W\left(\vec{\varphi}\left(t_k^{\text{bg}},\vec{X}_k^{\text{bg}};\theta\right);\lambda_{\text{bg}}\right)\right),
\end{align*}}
where the collocation points are sampled according to the partition $\Omega_0=\Omega_\text{myo}\dot{\cup}\text{ }\Omega_{\text{bg}}$, with $\{\vec{X}_k^\text{myo}\}_{k=1}^{N_\text{myo}}$ belonging to the myocardial tissue $\Omega_\text{myo}$ and $\{\vec{X}_k^\text{bg}\}_{k=1}^{N_\text{bg}}$ belonging to the background $\Omega_\text{bg}$. The function $W(\varphi;\lambda)$ is a neo-Hookean hyper-elastic strain energy function in three dimensional space that acts as a regularizer and has the form
\begin{align*}
    W(\varphi;\lambda) = {\rm Tr}(\mathbf{C}) - 3 - 2 \log(J)+\lambda(J-1)^2,
\end{align*}
with $\mathbf{C}:=({\text{d}\vec{\varphi}/\text{d}\vec{X}})^T(\text{d}\vec{\varphi}/\text{d}\vec{X})$ the right Cauchy-Green deformation tensor and $J:=\det (\text{d}\vec{\varphi}/\text{d}\vec{X})$ the determinant of the jacobian. Qualitatively, the determinant $J$ measures local changes in volume at each point of the material and the parameter $\lambda>0$ quantifies the material incompressibility. The term $\lambda(J-1)^2$ penalizes bulk variations of the medium. It is zero when volume is preserved ($J=1$) and strictly positive when the material suffers volumetric transformations ($J\neq 1$). Then, in order to enforce quasi-incompressibility within the myocardial region $\Omega_{\text{myo}}$, we assert that the bulk modulus is significantly greater in cardiac muscle than in the surrounding tissue, which we impose by setting $\lambda_{\text{myo}}\gg \lambda_\text{bg}$. This condition allows the background to deform freely while avoiding non-positive values for $J$ that could otherwise lead to non-diffeomorphic maps. 

\subsubsection{WarpPINN-fibers: Model Enhancement with Fiber Mechanics}\label{warpinnFibers}
During ventricular systole, cardiac muscle contracts to pump blood out of the ventricular chambers. This process is locally determined by the shortening of myocytes in response to a depolarization wavefront paced by the atrio-ventricular node and marked by the helical direction in which these cells are arranged \cite{Huxley1954,HuxleyHanson1954}.  
To incorporate this mechanical response into the WarpPINN model, we assume there exists a unit vector field $\mathbf{f}:\Omega_{\text{myo}}\to\mathbb{R}^3$ that encodes the direction of fibers at end-diastole. Then, for the deformation map $\vec{\varphi}(t,\vec{X})$, we measure fiber stretch $\lambda_{\mathbf{f}}^2$ by projecting the right Cauchy-Green deformation tensor of $\vec{\varphi}$ onto the fiber direction $\mathbf{f}$:
\begin{align*}
\lambda_\text{f}^2:=\mathbf{C}:\mathbf{f}\otimes\mathbf{f}
\end{align*}

The shortening of cardio-myocytes for an entire beat is then expressed by the condition $\lambda_\mathbf{f}^2<1$ for all times, while stretching is described by $\lambda_\mathbf{f}^2>1$. Since fiber stretching is unphysiological in healthy tissue, we aim to incorporate the mechanical constraint $\lambda_\mathbf{f}^2\leq 1$ into the WarpPINN model. To achieve this, we propose a fiber-informed regularizer of the form 
\begin{align*}
    \mu_\mathbf{f}\left|\left|\max\left\{1,\lambda_\mathbf{f}^2(\vec{\varphi})\right\} - 1\right|\right|^q_q
\end{align*}
Note that the function $\max\{1,\lambda\} - 1$ outputs zero when $\lambda\leq 1$ and equals $\lambda-1>0$ when $\lambda>1$. Therefore, the proposed regularizer promotes myocytic contraction and penalizes fiber stretching. 

For a set of collocation points $\{(t_k^\mathbf{f},\vec{X}_k^\mathbf{f})\}_{k=1}^{N_\mathbf{f}}\subseteq [0,1]\times\Omega_\text{myo}$ where fiber orientation is available at end-diastole, we postulate the following additional term to the loss function:
\begin{align*}
    \mathcal{L}_\mathbf{f}(\theta):=\mu_\mathbf{f}\dfrac{1}{N_\mathbf{f}}\displaystyle\sum\limits_{k=1}^{N_\mathbf{f}}\left(\max\left\{1,\lambda_\mathbf{f}^2\left(\vec{\varphi}(t_k^\mathbf{f},\vec{X}_{k}^\mathbf{f};\theta),\mathbf{f}(\vec{X}_k^\mathbf{f})\right)\right\}-1\right)^2,
\end{align*}
where we have considered $q=2$. Finally, we introduce WarpPINN-fibers as a fully-connected neural network architecture $\hat{\vec{u}}(t,\vec{X};\theta)$ with loss function
\begin{align*}
    \mathcal{L}(\theta;\mu,\lambda_\text{myo},\lambda_\text{bg},\mu_\mathbf{f}):=\mathcal{L}_0(\theta;\mu,\lambda_\text{myo},\lambda_\text{bg})+\mathcal{L}_\mathbf{f}(\theta;\mu_\mathbf{f})
\end{align*}

\begin{figure}[t]
	\centering
	\includegraphics[width=\textwidth]{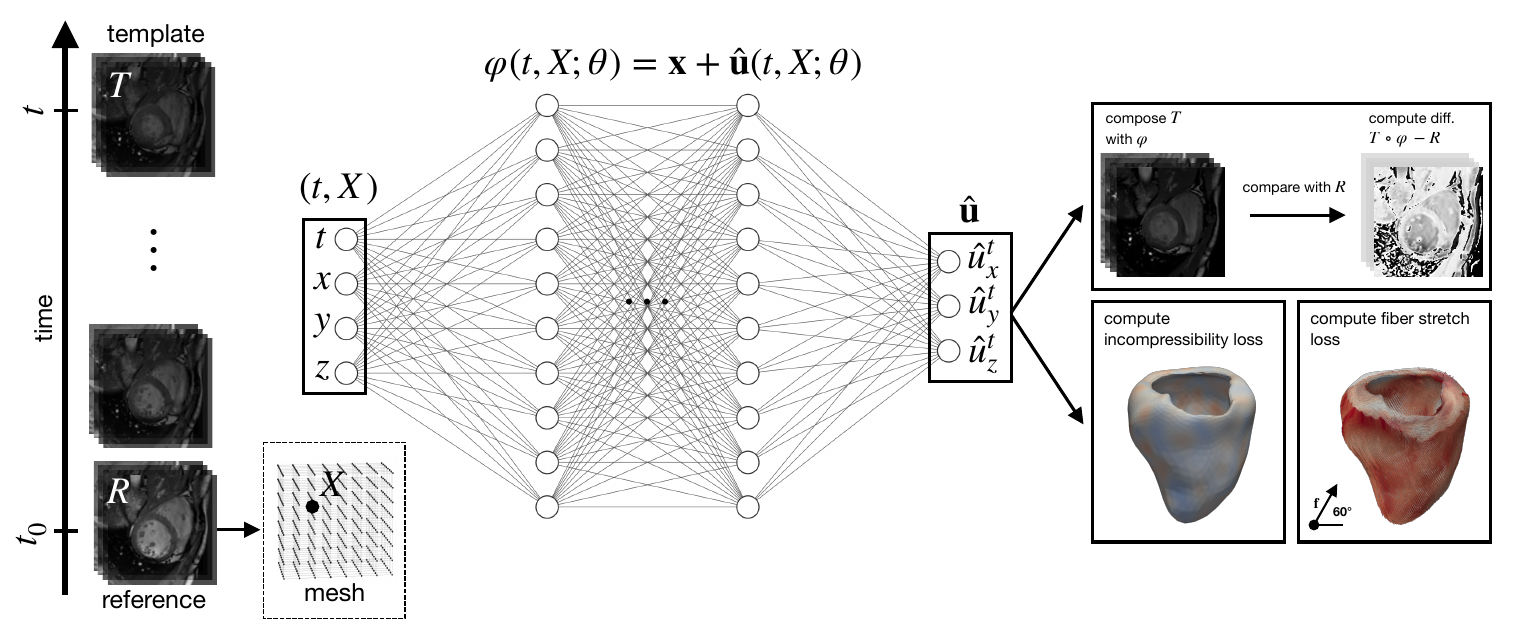}
	\caption{In the neural network, the inputs are the time $t$ and the spatial coordinates $\vec{X}$ of the reference image, and the output is the displacement field $\hat{\vec{u}}$. The objective is to find the displacement field that warps the template images to the end-diastolic state across the entire cardiac cycle and fulfills the quasi-incompressibility condition under the constraint of fiber stretch.}
	\label{fig:network}
\end{figure}

\subsection{Generation of Fibers with Rule-Based Method}
We employ computationally generated fibers to incorporate tissue anisotropy into our model without acquiring any additional data. In this study, we implement the Laplace-Dirichlet rule-based (LDRB) algorithm proposed in \cite{Bayer2012}, but other algorithms could be used, such as purely mathematical models \cite{BARNAFI2025117710} or other rule-based strategies \cite{wong2012, Doste2019,perotti2021estimating}.

For a biventricular heart domain, the LDRB algorithm creates a $3\times3$ matrix field $Q$ representing a fiber-oriented orthonormal coordinate system from a set of prescribed angle functions that describe the helical and sheet angles of the fibers. The LDRB method computes a set of gradient fields $\{\nabla\psi_\text{ab},\nabla\psi_\text{lv},\nabla\psi_\text{rv},\nabla\psi_\text{epi}\}$ in $\mathbb{R}^3$, each derived from a harmonic potential with a specific boundary condition. The gradient $\nabla\psi_\text{ab}$ defines the apico-basal direction, while $\nabla\psi_\text{lv}$ and $\nabla\psi_\text{rv}$ define transmural fields oriented from the endocardium to the epicardium in the left ventricle and the right ventricle, respectively. The direction $\nabla\psi_\text{epi}$ corresponds to a transmural field oriented from epicardium to endocardium. The coordinate system $Q$ is then constructed from a weighted averaging procedure (bislerp) over a set of $3\times3$ coordinate frames $\{Q_\text{lv},Q_\text{rv},Q_\text{epi}\}$, where each frame $Q_\text{d}$ is obtained by rotating the resulting matrix from the Gram–Schmidt orthonormalization of $\{\nabla\psi_\text{d},\nabla\psi_\text{ab}\}$ according to the provided angle functions. 

The helical angle functions determine the rotation about the radial axis for each frame. They are characterized by user-specified angles $\alpha_\text{endo}$ and $\alpha_\text{epi}$ for the endocardial and epicardial surfaces, respectively. Similarly, the sheet angle functions define the rotation about the circumferential component and are defined by the angles $\beta_\text{endo}$ and $\beta_\text{epi}$, which are also specified by the user. 

In our setting, we restrict to the particular case $\alpha_\text{endo}=-\alpha_\text{epi}=:\alpha$, $\beta_\text{endo}=\beta_\text{epi}=0$ and use linear interpolation to construct the angle functions \cite{Bayer2012}. Then, the only hyper-parameter required to construct the fibers is the helical angle $\alpha\in[-\pi,\pi]$.  


\subsection{Fourier Feature Mapping and Weights Initialization}
In the context of deep learning, Fourier feature mappings (FFMs) transform low-frequency inputs to high-frequency features in a space of increased dimension to overcome the problem of spectral bias during training \cite{rahaman19a,wang2021eigenvector,tancik2020fourier}. 
Let $\mathcal{X}$ be an input space in $\mathbb{R}^n$. For an integer $m>0$, that determines the dimensionality of the new feature space, and a real value $\sigma>0$, that controls the frequency spectrum of the input domain, we compute a matrix $\mathbf{B}\in\mathbb{R}^{m\times n}$, with entries independently drawn from a normal distribution $\mathcal{N}(0,\sigma^2)$, and define the FFM as the function
\begin{samepage}
\begin{align*}
    \gamma:&\mathcal{X}\to\mathbb{R}^{2m}\\
    &\vec{X}\mapsto     \gamma(\vec{X})=\begin{bmatrix}
        \cos \left(\mathbf{B}\vec{X}\right) \\
        \sin \left(\mathbf{B}\vec{X}\right)
    \end{bmatrix}
\end{align*}
\end{samepage}
When equipped with Fourier features for the spatial coordinate, the WarpPINN model significantly improves performance in the reconstruction of cardiac strains \cite{lopez2023warppinn}. We adopt the same methodology for input processing in the WarpPINN-fibers architecture.

To prevent our neural network $\hat{\vec{u}}$ from returning displacement fields that produce negative volumes, we perform a pre-training step to approximate the identity map. First, we initialize the weights of our neural network with  Xavier initialization and set bias vectors to $0$. Then, we train our neural network with the Adam optimizer \cite{kingma2014adam} to minimize the loss function
\begin{align*}
    \mathcal{L}_\text{pre}(\theta):=\dfrac{1}{N_\text{fr}}\displaystyle\sum\limits_{i=1}^{N_\text{fr}}\dfrac{1}{N_d}\sum\limits_{j=1}^{N_\text{d}}\left|\left|\hat{\vec{u}}(t_i,\vec{X}_j^\text{d};\theta)\right|\right|_2^2 
\end{align*}

This way, we predict near-null displacements in the entire domain and, subsequently, get the estimate $\vec{\varphi}(t,\vec{X};\theta)=\vec{X}+\hat{\vec{u}}(t,\vec{X};\theta)\simeq \vec{X}$ for the resulting weights $\theta$.
\subsection{Training of the Network}
We train the WarpPINN-fibers architecture with the Adam optimizer to minimize the loss $\mathcal{L}(\theta)$ presented in Section~\ref{warpinnFibers}. This function is made up of a data similarity term, a neo-Hookean regularization and a fiber-stretch penalization:

{\small\begin{align*}
    \mathcal{L}(\theta)=&\dfrac{1}{N_\text{fr}}\displaystyle\sum\limits_{i=1}^{N_\text{fr}}\dfrac{1}{N_\text{d}}\sum\limits_{j=1}^{N_\text{d}}\left|R(\vec{X}_j^\text{d})-T_{t_i}(\vec{\varphi}(t_i,\vec{X}_j^\text{d};\theta))\right|\\
    &+\mu\left(\dfrac{1}{N_\text{myo}}\displaystyle\sum\limits_{k=1}^{N_\text{myo}}W\left(\vec{\varphi}\left(t_k^{\text{myo}},\vec{X}_k^{\text{myo}};\theta\right);\lambda_{\text{myo}}\right)+\dfrac{1}{N_\text{bg}}\displaystyle\sum\limits_{k=1}^{N_\text{bg}}W\left(\vec{\varphi}\left(t_k^{\text{bg}},\vec{X}_k^{\text{bg}};\theta\right);\lambda_{\text{bg}}\right)\right)\\
    &+ \mu_{\mathbf{f}} \frac{1}{N_{\mathbf{f}}} \sum_{k=1}^{N_{\mathbf{f}}} \left(\max \left\{1,\mathbf{C}({\vec{\varphi}}(t_k^\mathbf{f},\vec{X}_k^{\mathbf{f}}; {\theta})):\mathbf{f}(\vec{X}_k^{\mathbf{f}}) \otimes \mathbf{f}(\vec{X}_k^{\mathbf{f}})\right\}-1\right)^2.
\end{align*}}

The first term is computationally expensive to evaluate due to the trilinear interpolation required for the composition of $T_{t_i}$ with the deformation field $\vec{\varphi}$. To address this problem, at each iteration we fix a time $t\in\{t_i\}_{i=1}^{N_\text{fr}}$ and only compute the difference between $R$ and $T_{t}\circ\vec{\varphi}$ at that particular time. The time $t$ is selected sequentially starting from $t_1$, stepping from $t_i$ to $t_{i+1}$ at the end of each training iteration and cycling back to $t_1$ once all times are used.
For the second term, we use mini-batches of time-space for the collocation points $\{(t_k^\text{myo},\vec{X}_k^{\text{myo}})\}_{k=1}^{N_\text{myo}} \subset [0,1]\times\Omega_{\text{myo}}$ and $\{(t_k^\text{bg},\vec{X}_k^{\text{bg}})\}_{k=1}^{N_\text{bg}} \subset [0,1]\times\Omega_{\text{bg}}$, where only the times $\{t_k^\text{myo}\}_{k=1}^{N_\text{myo}}\cup\{t_k^\text{bg}\}_{k=1}^{N_\text{bg}}$ are randomly resampled at each training iteration.
For the last term, we have a set of points where the fibers orientation is defined at end-diastole, so at each iteration we randomly select a subset $\{\vec{X}_k^\mathbf{f}\}_{k=1}^{N_\mathbf{f}}$ of these points, pair them with uniformly sampled times $\{t_k\}_{k=1}^{N_\mathbf{f}}\subseteq[0,1]$ and compute the fiber stretch. 

\subsection{Synthetic Phantom Experiment}

To evaluate the performance of the WarpPINN-fibers approach, we first test our methodology with the analytical phantom from \cite{perotti2021estimating} resembling the middle region of the left ventricle. The phantom consists of a thick-walled cylinder model with fibers that is subjected to a known compressive and torsional deformation field $\mathbf{d}$. In this model, cardio-myocyte orientations are defined by rotations applied to the circumferential and transmural directions of the geometry based on histological observations from excised ovine hearts \cite{perotti2021estimating,ennis2008myofiber}.
The phantom geometry is centered along the z-axis and is described in cylindrical coordinates $(r,\tau,z)\in[R_\text{endo},R_\text{epi}]\times[-\pi,\pi)\times[Z_\text{bott},Z_\text{top}]$, with $R_\text{endo}=20$, $R_\text{epi}=35$ and $-Z_\text{bott}=Z_\text{top}=10$. We set the helix angles to $37$° on the endocardium, $-9$° on the ventricular mid-wall and $-45$° on the epicardium.

The deformation field applied to this model is also expressed in cylindrical coordinates, in the form $\mathbf{d}(t,r,\tau,z)=(d_{\hat{r}}(t,r,\tau,z),d_{\hat{\tau}}(t,r,\tau,z),d_{\hat{z}}(t,r,\tau,z))^T$, and is defined as

\begin{equation}
    \begin{aligned}
    {d}_{\hat{r}}&:= s(t)\cdot(a_0 + a_1 r + a_2 r^2), \\
    {d}_{\hat{\tau}} &:= s(t)\cdot a_3\frac{z - Z_{\text{bott}}}{Z_{\text{top}} - Z_{\text{bott}}}, \\
    {d}_{\hat{z}} &:= z + s(t)\cdot a_4 z,
    \end{aligned}
\end{equation}

Here, the parameters $\{a_l\}_{l=0}^4\subseteq \mathbb{R}$ are adjustable and serve to prescribe torsion, compression and incompressibility of the cylinder. The function $s(t)$ is a periodic map of the form $\sin^2(\pi t)$ defined over the time interval $[0,1]$ that scales each $a_j$ to emulate systolic motion as $t$ approaches $1/2$. For simplicity, we focus in studying a static example.

We fix $a_0 = -16.95$, $a_1 = 0.76$, $a_2 = -0.01$, $a_3 = 2.54$, $a_4 = -0.11$ and bound the space of observation to a rectangular prism $\Omega_0:=[-40,40]\times[-40,40]\times[-10,10]$ that contains the phantom. 

To obtain the reference image $R$, we take a snapshot at time $t=0$, when the cylinder is at rest, assigning $R(\vec{X})=0$ to the background and $R(\vec{X})=f(\vec{X})$ to the cylinder, with  
\begin{align*}
       f(x,y,z):=\sin\left(4\pi \left(\frac{x}{40}+ \frac{z}{10}\right)\right)  \cdot \cos \left(4 \pi \left(\frac{y}{40}+ \frac{z}{10}\right)\right) 
\end{align*}
After creating the reference image $R$, we take another snapshot at the intermediate time $t=4/9$, when the cylinder is partially deformed. Then, we define the pixel intensity $T(\vec{X})=R\left({\mathbf{d}}^{-1}(\vec{X};t=4/9)\right)$ for each point in the deformed cylinder, while assigning uniform intensity $T(\vec{X})=0$ elsewhere. Note that in this setting we only add texture to the phantom, disregarding any displacement information from the background.
Finally, we add Gaussian noise with amplitude $\sigma =10^{-4}$ to both the reference and the template (see Figure~\ref{app:synth_images}).


Since we are only interested in recovering the deformation map $\mathbf{d}$ from one template image $T$, we consider a WarpPINN-fibers architecture $\hat{\vec{u}}(\vec{X};\theta)$ that is independent of time. Temporal dependency could be included to create a field smoothly varying in time to transition from the reference to the template. However, this could be counterproductive, since the warped geometry at interpolated times would not be attributable to observed images of the object. The architecture $\hat{\vec{u}}(\vec{X};\theta)$ consists of $5$ hidden layers, each containing $64$ neurons.
We use Fourier features with $m = 8$ and $\sigma = 1$, resulting in an input layer of dimension $2m = 16$ and an output layer of dimension $3$.
The neo-Hookean strain energy function $W$ is evaluated at $20,000$ collocation points, which contribute to each term of the quasi-incompressibility loss based on whether they belong to the cylinder. The fiber stretch penalization is evaluated over another set of collocation points of size $1,000$. Collocation points are randomly sampled within the space of the reference image at each iteration.

The strain energy for the background is calculated using $\lambda_{\text{bg}} = 50$, while the strain energy for the cylinder is computed with $\lambda_{\text{myo}} = 10^4$, which enforces the quasi-incompressibility in this region.
The selection of a value for $\lambda_{\text{bg}}$ is justified by the fact that the background has no texture and should deform freely, but constrained to physical motion.

We also set $\mu = 10^{-2}$ and $\mu_{\mathbf{f}} = 100$.
The pre-training phase is conducted until the displacement loss $\mathcal{L}_\text{pre}(\theta)$ reaches a value below the threshold $10^{-6}$. After this, the network is trained for the global loss $\mathcal{L}(\theta)$ across $10^{4}$ iterations.

To evaluate the performance of WarpPINN-fibers in this synthetic example, we compute the mean square error between the displacement $\mathbf{u}(\vec{X})=\mathbf{d}(\vec{X};t=4/9)-\mathbf{d}(\vec{X};t=0)$ associated to the deformation field $\mathbf{d}$ and the predicted displacement field $\hat{\vec{u}}(\vec{X};\theta)$:
\begin{align*}
    \displaystyle\frac{1}{N_\text{d}} \sum_{j=1} \left|\left|\mathbf{u}(\vec{X}_j^\text{d}) - \hat{\vec{u}}(\vec{X}_j^\text{d};\theta)\right|\right|^{2},
\end{align*}
and compare these results with the mean square error produced by the WarpPINN model.
\subsection{Application on MICCAI Cine-MRI Dataset}
    To validate our methodology in a realistic clinical framework for the MR image registration problem, we apply the warpPINN-fibers approach to cine steady-state free precession (SSFP) MRI from a public dataset available through The Cardiac Atlas Project\footnote{https://www.cardiacatlas.org/motion-tracking-2011-challenge/}. This dataset was presented at the 2011 MICCAI workshop ``Statistical Atlases and Computational Models of the Heart: Imaging and Modelling Challenges’’ (STACOM-11) for a motion tracking problem of the left ventricle \cite{stacom2011}. The STACOM-11 dataset contains cine SSFP MR images with 12 manually tracked landmarks from 15 healthy volunteers, alongside a left ventricular surface mesh at end-diastole for each volunteer. 
    
    Unlike the synthetic experiment performed in the previous section, we now have access to a sequence of images $\{T_{t_i}\}_{i=0}^{N_\text{fr}}$ ($R:=T_{t_0}$) capturing different stages of the cardiac cycle from end-diastole to end-diastole. We only consider snapshots from the short axis. However, other axes could be included in our pipeline, such as 2-chamber and 4-chamber views. For more details regarding image resolution and number of frames available for each patient, we refer the reader to \cite{stacom2011}.


The end-diastolic left ventricular mesh allows to define the myocardium points $\{\vec{X}_k^{\text{myo}}\}_{k=1}^{N_{\text{myo}}}$ and the background points $\{\vec{X}_k^{\text{bg}}\}_{k=1}^{N_{\text{bg}}}$ for the image domain of each volunteer. Additionally, from each mesh we can sample the collocation points $\{\vec{X}_k^\mathbf{f}\}_{k=1}^{N_\mathbf{f}}$ for the fiber-stretch penalization. 


We compute the fibers orientation in a refined volumetric mesh of each cardiac geometry with the LDRB algorithm \citep{Bayer2012} for a specified helix angle $\alpha$ and the sheet angle $\beta$ fixed at $0^\circ$. Since real fiber orientation is unknown, we cannot verify if a chosen angle $\alpha$ accurately reflects the anatomical fiber distribution of the patient. However, we can analyze the impact of different values of $\alpha$ to the WarpPINN-fibers predictions and the influence of fiber stretch to the network outputs. For this purpose, we compare the accuracy of our method with different fiber angles $\alpha\in\{50\text{°},60\text{°},70\text{°}\}$ and varying weights $\mu_\mathbf{f}\in\left\{10^{i}\right\}_{i=-3}^3$ for the fibers loss.

Fibers computed at the base of each mesh tended to be disorganized and in disagreement with the helical pattern expected, possibly due to the base not being defined by a horizontal cut, as in the examples of the algorithm. To mitigate the effects of chaotic fiber distribution of the points at the base, we exclude them from the fiber-stretch penalization. 

We compare the performance of our WarpPINN-fibers architecture with four algorithms that have been previously applied to the STACOM-2011 dataset: Temporal Diffeomorphic Free Form Deformation, iLogDemons, CarMEN and WarpPINN. The first two methods were used as benchmarks for this dataset \cite{stacom2011}. 

The Temporal Diffeomorphic Free Form Deformation approach was developed by researchers from Universitat Pompeau Fabra (UPF) \cite{de2012temporal} and extends the classical registration framework by describing cardiac motion through a velocity field from which displacement is computed via integration. In the UPF method, velocity is presented as a sum of continuous spatiotemporal B-spline kernels. This approach estimates a similarity metric that compares each image to the reference image, as well as frame-to-frame similarities, and incorporates a quasi-incompressibility condition with a regularizer that enforces null-divergence (solenoidal) velocity fields. iLogDemons was proposed by the Inria-Asclepios project (INRIA) \cite{mansi2011ilogdemons} and works under a similar setting. This algorithm processes each frame independently, with a stationary velocity field and an additional divergence-free constraint to ensure quasi-incompressibility. The third method, CarMEN \cite{morales2021deepstrain}, uses a convolutional neural network architecture for myocardial strain analysis, receiving a reference image and a template image as inputs to estimate the associated motion from the reference. The last method, WarpPINN, is a physics-informed neural network and is detailed in section \ref{warpinn}


The STACOM-2011 dataset provides a set of 12 ground-truth landmarks for each volunteer to assess cardiac strain estimation: one landmark per wall (anterior, lateral, posterior and septal) for each ventricular level (basal, mid-ventricular and apical). These landmarks were manually tracked by two observers from a 3D tagged MR dataset over the cardiac cycle \cite{stacom2011} and then converted to SSFP coordinates. Due to temporal misalignment during conversion to SSFP, landmarks can only be spatially mapped to end-systole (intermediate frame) and to end-diastole (last frame). Therefore we evaluate predicted displacements only at landmark points on these two states: For a trained WarpPINN-fibers predictor, the ground-truth landmarks from the first frame are used as inputs to predict their position at subsequent frames up until the final frame at $t=1$. We measure the error for each method with the Euclidean difference between the predicted and the ground-truth positions from the end-systolic frame and the end-diastolic final frame. 


In addition to landmark tracking, we measure the average values for the longitudinal, circumferential and radial components of the Lagrange strain induced by WarpPINN-fibers over a selected set of AHA segments \cite{SELVADURAI2018926} at each step of the cardiac cycle. Since there are no ground-truth strain curves for direct evaluation in this case, we compare our results against the UPF, INRIA and WarpPINN approaches.

For a given deformation field $\vec{\varphi}(t,\vec{X})$, the Lagrange strain tensor is defined as
\begin{align*}
    \mathbf{E}(t,\vec{X}) :=& \frac{1}{2} (\mathbf{C}(\vec{\varphi}(t,\vec{X})) - \mathbf{I}),
\end{align*}

where $\mathbf{I}$ denotes the identity map in $\mathbb{R}^3$. We decompose the Lagrange strain tensor $\mathbf{E}$ according to a local coordinate system $\mathbf{P}=(\mathbf{l}\text{ } \mathbf{r} \text{ } \mathbf{c})$ based on the anatomy of the left ventricle, consisting of a longitudinal direction $\mathbf{l}$, a radial direction $\mathbf{r}$ and a circumferential direction $\mathbf{c}$:
\begin{align*}
    \mathbf{E}&=\mathbf{E}_{\mathbf{ll}}(\mathbf{l}\otimes\mathbf{l})+\mathbf{E}_{\mathbf{rr}}(\mathbf{r}\otimes\mathbf{r})+\mathbf{E}_{\mathbf{cc}}(\mathbf{c}\otimes \mathbf{c})\\
    &:=\left[\mathbf{l}^T\mathbf{E}\mathbf{l}\right](\mathbf{l}\otimes \mathbf{l})+\left[\mathbf{r}^T\mathbf{E}\mathbf{r}\right](\mathbf{r}\otimes \mathbf{r})+\left[\mathbf{c}^T\mathbf{E}\mathbf{c}\right](\mathbf{c}\otimes \mathbf{c}),
\end{align*}

where we evaluate the strains $\mathbf{E}_\mathbf{ll}$, $\mathbf{E}_\mathbf{rr}$ and $\mathbf{E}_\mathbf{cc}$ on the ventricular surface. The field $\mathbf{l}$ is oriented in a straight line drawn from the apex to the center of the mitral valve. The meshes from the dataset are all aligned with this direction, so we have $\mathbf{l} := (0,0,1)^T$ uniformly on each point of the mesh. The radial direction $\mathbf{r}$ is obtained via orthogonalization of the normal vector field $\mathbf{n}$ with respect to $\mathbf{l}$ and then normalized, so $\mathbf{r} := (\mathbf{n} - (\mathbf{n} \cdot \mathbf{l})\mathbf{l})/||\mathbf{n} - (\mathbf{n} \cdot \mathbf{l})\mathbf{l}||$. Finally, the circumferential direction $\mathbf{c}$ is given by the cross product of the two vector fields $\mathbf{l}$ and $\mathbf{r}$, that is $\mathbf{c} := \mathbf{l} \times \mathbf{r}$ \cite{stacom2011}.

For strain curve visualization, we use several subsets from the 17 American Heart Association (AHA) segments covering the basal, mid-cavity and apical regions of the left ventricle, following the guidelines from \cite{stacom2011}. For each anatomy, we compute the average strain at each time on each of these three sections.

The network is configured with $5$ hidden layers, having $64$ neurons per layer. For the FF mapping, we set $m = 32$ and $\sigma = 1$, yielding $64$ spatial features and a single variable for time, resulting in an input dimension of $2m + 1 = 65$ and an output dimension $3$. We evaluate the neo-Hookean energy with $\lambda_{\text{inc}} = 10^{5}$ and $\lambda_{\text{bg}} = 1$ at $2,000$ collocation sites per iteration, $1,000$ within the ventricular region and $1,000$ lying in the background. The fiber regularizer is computed over the same set of $1,000$ collocation points employed to evaluate $W$ within the ventricular region.

Pre-training is performed with a convergence tolerance of $10^{-5}$. Following pre-training, our network is trained for approximately $3,500$ epochs without the fiber regularizer to generate a baseline estimate for the displacement solution. Then, we consider two training scenarios: 
\begin{enumerate}
    \item WarpPINN: The network is subsequently trained for another $1,500$ epochs with no fiber regularizer.
    \item WarpPINN-fibers: The network is subsequently trained for another $1,500$ epochs employing the fiber regularizer. 
\end{enumerate}
This strategy allows to establish a fair comparison between the WarpPINN and WarpPINN-fibers model, since the architecture and hyper-parameters of the network are preserved, while the physical model defining each method changes.




\section{Results}
\subsection{Synthetic Phantom Experiment}
We assess the effectiveness of the WarpPINN-fibers model to estimate anisotropic motion in our synthetic example by computing the mean square error (MSE) relative to a known displacement field and measuring fiber-stretch across the geometry. We compare these results with the WarpPINN model outputs to elucidate the predictive differences between sole quasi-incompressibility and quasi-incompressibility complemented with fiber mechanics.  

\begin{figure}[t]
	\centering
	\includegraphics[width=\textwidth]{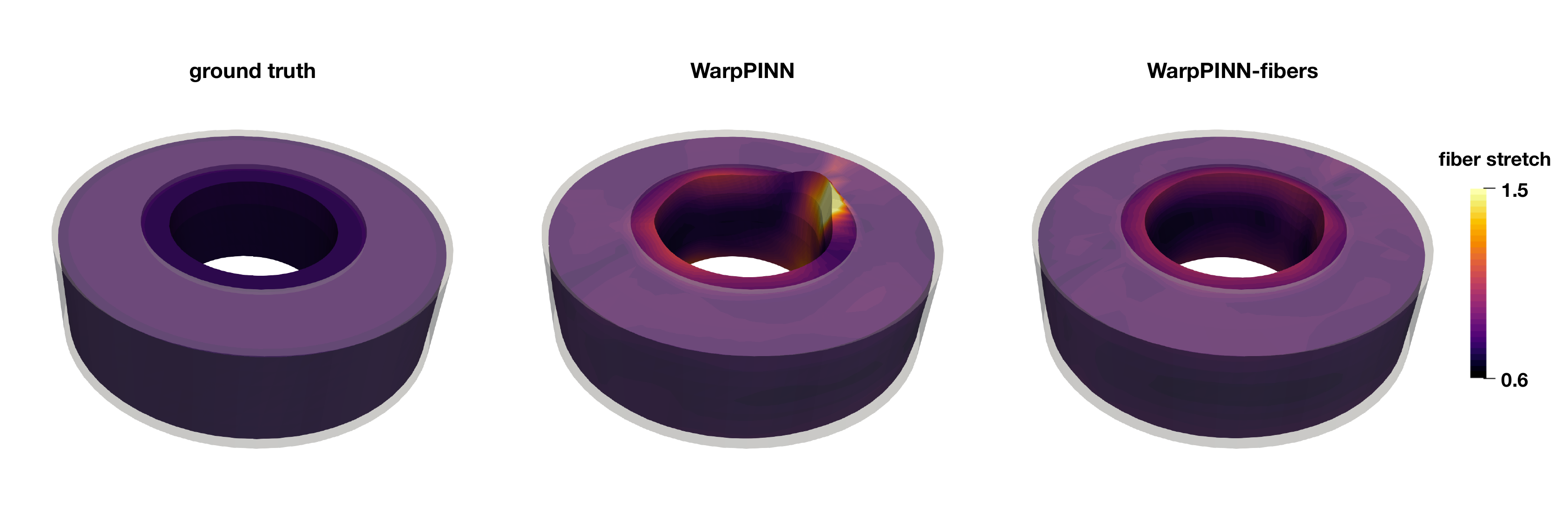}
	\caption{Fiber stretch over the cylindrical phantom for the ground truth (left), WarpPINN (center) and WarpPINN-fibers (right). We show the deformed configuration of the object in three dimensions overlaid with a translucent view of the resting state on each case.}
	\label{fig:synth_fiber_stretch}
\end{figure}

We report a MSE of $1.16 \cdot 10^{-1}$ for WarpPINN and $6.47 \cdot 10^{-2}$ for WarpPINN-fibers. Similar values are obtained for the second method equipped with FF mappings with a moderate to low frequency, as seen in ~\ref{app:FF_sigma_m}. The reported MSE values indicate that the global deformation derived from both models is in agreement with the ground-truth data, displaying the fibers-based model slight improvement in comparison to the other approach. However, a substantial difference in local displacement and torsion are observed in Figure \ref{fig:synth_fiber_stretch}. Here, we show the deformed configurations of the cylindrical phantom for the ground-truth displacement, the displacement predicted by WarpPINN and the displacement predicted by WarpPINN-fibers, with the associated fiber-stretch encoded by a linear colormap. The analytical deformation displays a smooth and homogeneous behavior, with a uniformly distributed fiber stretch $\lambda_\mathbf{f}^2$ below $1$ across the entire geometry. WarpPINN replicates this behavior in most parts of the domain, except at the top end of the shape, where the cylinder appears to be stretching out in a specific region of the inner ring and exhibits an increased fiber elongation within that region, reaching values that exceed $1$. WarpPINN-fibers fixes this issue, predicting a homogeneous stretch similar to the ground truth and a deformation field that matches the real warping of the cylinder albeit small differences on the edges of the inner annuls.

\begin{figure}[t]
	\centering
	\includegraphics[width=\textwidth]{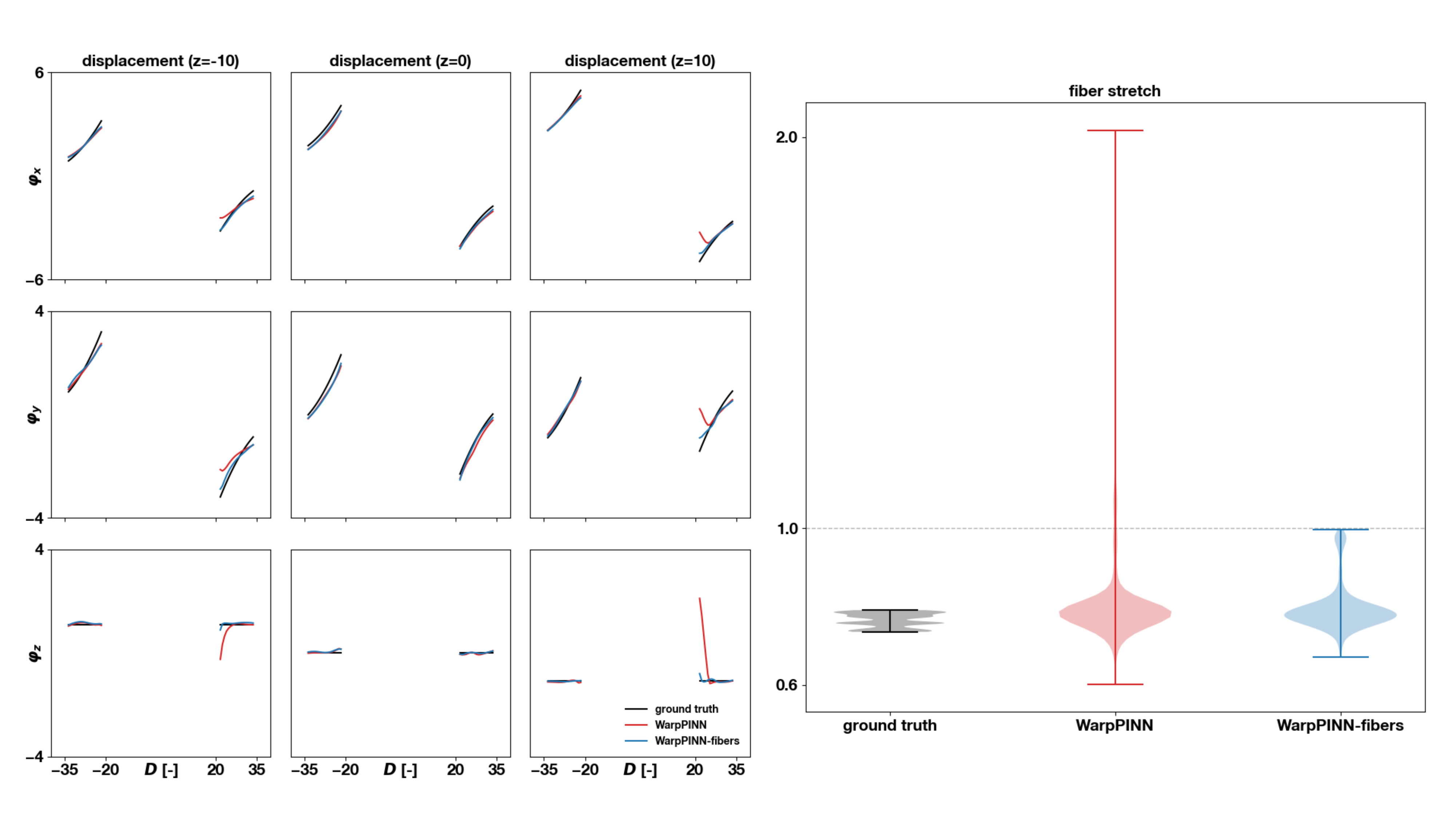}
	\caption{Displacement and fiber stretch of synthetic example for the ground truth (black), WarpPINN (red) and WarpPINN-fibers (blue). On the left, we measure each component of the displacement field for diagonal profiles $\{x=y\}$ at the bottom, ($z=-10$), mid-section ($z=0$) and top ($z=10$) of the cylinder. On the right, we show the violin plots of the fiber stretch over the entire cylinder for the ground truth, WarpPINN and WarpPINN-fibers.}
	\label{fig:synth_disp_strain_fib}
\end{figure}

The improvement in performance introduced by WarpPINN-fibers is further reinforced with the plots from Figure \ref{fig:synth_disp_strain_fib}. On the left we exhibit the $x$-,$y$- and $z$- directions of the predicted displacement fields and components at a diagonal profile $\{x=y\}$ in the bottom cap ($z=-10$), in a middle cut ($z=0$) and in the top cap ($z=10$) of the cylinder. Here, we observe that both the WarpPINN and WarpPINN-fibers predictions lie within the analytical displacement components in the middle region and in the points farther from the extrusion zone observed in Figure~\ref{fig:synth_fiber_stretch}, presenting small variations from the ground-truth profiles. However, WarpPINN diverges from the solution in the top and bottom boundaries of the cylinder where the extrusion occurs, reaching maximum errors of approximately $1.7$, $1.6$ and $3.2$ on the $x$, $y$ and $z$ coordinates, respectively. In contrast, WarpPINN-fibers reaches errors of at most $0.5$ in the same set of directions. 

 From the violin plot from Figure~\ref{fig:synth_disp_strain_fib}, we can see that, although WarpPINN presents a global fiber stretch ($0.81$ on avg.) close to the reference data ($0.76$ on avg.), it overestimates lengthening and surpasses the range of ground-truth measurements, reaching stretching as high as $2$ and as low as $0.6$. On the other hand, WarpPINN-fibers shows a lower fiber lengthening compared to the other method ($0.79$ on avg.) and displays smaller deviations from the average, remaining upper-bounded by $1$. This indicates that the entire geometry is contracting with our fiber-based estimator, which is not the case for the other method. We additionally inspect the agreement of the tested models to warp the textured image in a horizontal cut of the cylinder at $z=0$ (see \ref{app:synth_images}). In this case, we observe that both methods produce images that resemble the real snapshot.

\subsection{MICCAI Cine-MRI Dataset}

We validate WarpPINN-fibers for the 3D+t image registration task using the cine SSFP MR data derived from 15 subjects in the STACOM-2011 dataset. Training the neural network on a single cluster node with a NVIDIA QUADRO RTX 8000 GPU took approximately 140 minutes for each volunteer. In what follows, we use the format ``vx’’ to refer to ``volunteer x’’. 

\begin{figure}[t]
	\centering
	\includegraphics[width=\textwidth]{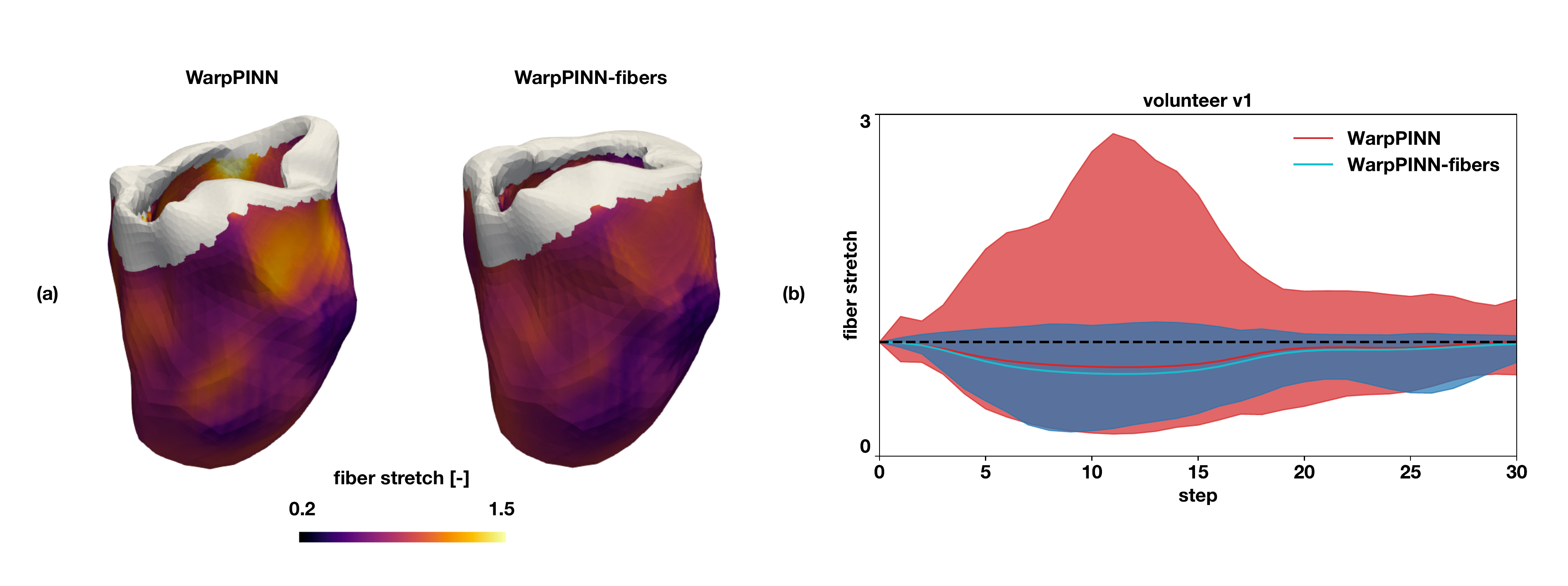}
	\caption{Influence of fiber penalization on deformation prediction and fiber stretch.
    (a) Fiber stretch in the deformed configuration of v1 at end-systole for WarpPINN (left) and WarpPINN-fibers (right). The base of the ventricle is shown in gray since fiber stretch was not properly computed in this region due to fiber misalignment.
    (b) Mean curves with min-max range of fibers stretch for WarpPINN (red) and WarpPINN-fibers (blue) at each step of the cardiac cycle. The stretch at the base of the ventricles are not considered.}
	\label{fig:fiber_stretch}
\end{figure}

First, we compare the fiber-stretch induced by the WarpPINN and the WarpPINN-fibers models on v1 across the entire cardiac cycle. In Figure~\ref{fig:fiber_stretch}, we show the deformed configuration of v1 obtained with both architectures at end-systole and display the corresponding fiber stretch at each point, excluding the nodes from the base. We observe that both models produce similar patterns of fiber stretch, but with different magnitudes, with WarpPINN exhibiting more prominent stretching in certain regions compared to our network. We also identify some differences between the deformation field predicted with our fiber-informed approach and the one predicted with the other method: In the base of the ventricle, WarpPINN produces a pronounced undulation of the mitral annulus and elongates tissue in the vertical axis, while WarpPINN-fibers predicts a uniform contraction at the base without longitudinal elongation.
In the same figure, we present a plot for fiber stretch versus time. It can be noticed that there is an overall tendency of the myocardial tissue to contract during the cardiac cycle, as reflected by the fact that the mean curves and minimum values for the assessed methods remain consistently below $1$. These curves decrease smoothly from diastole until a peak systolic state is reached; then, they increase and get closer to $1$ during myocardial relaxation. Additionally, we see that the fiber strains predicted using WarpPINN notably surpass the stretching threshold, showing maximum values that continuously weight in favor of elongation and minima that exaggerate fiber shortening. In contrast, the fiber stretch obtained with WarpPINN-fibers evolves in a more controlled manner, exceeding $1$ in some cases just by a slight amount. We disclose these pathological cases for WarpPINN-fibers occur near the basal region of the ventricles, where some fiber misalignment is produced.


\begin{figure}[t]
	\centering
	\includegraphics[width=\textwidth]{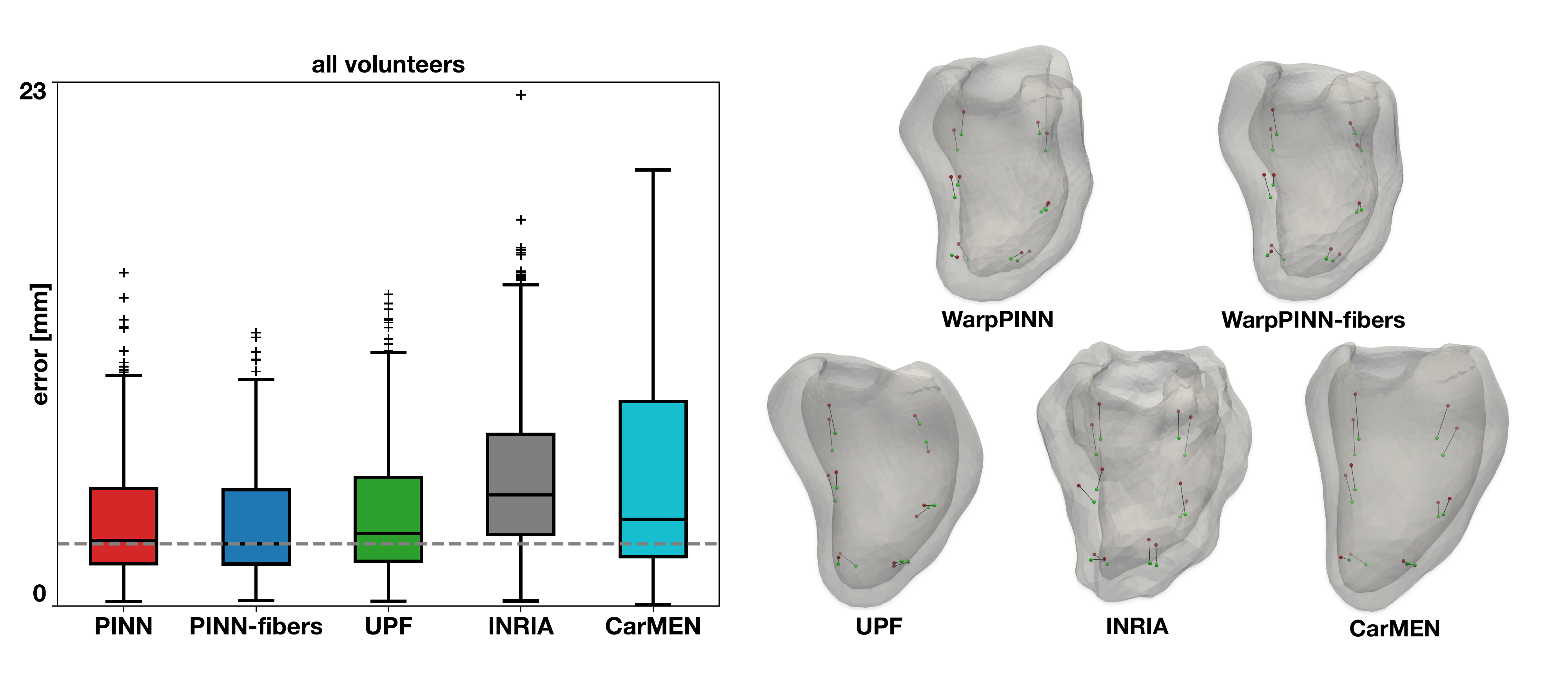}
	\caption{Predictions for WarpPINN (PINN), WarpPINN-fibers (PINN-fibers), UPF, INRIA and Carmen for manually-tracked landmarks. On the left, we show the box plots for landmark tracking error measured in millimeters. The dotted gray line indicates the lowest median achieved, which is obtained with WarpPINN-fibers for the parameters $\mu = 10^{-5}$, $\sigma = 1$, and $\mu_{\mathbf{f}} = 10$. On the right, we show the deformed configuration of v1 at end-systole for each method with the predicted landmarks (red) and ground-truth landmarks (green). Each predicted landmark is connected to its ground-truth location with a black line.}
	\label{fig:boxplot_mesh_lmks}
\end{figure}

Figure \ref{fig:boxplot_mesh_lmks} presents box plots showing the distribution of the errors between the manually-tracked landmarks and the respective positions predicted by the different models.
The median error for the WarpPINN architecture trained with $\mu = 10^{-5}$ and $\sigma = 1$ is approximately \SI{2.87}{[\m\metre]}, while our fiber-based approach trained with the same parameters and $\mu_{\rm{fib}} = 10$ exhibits less variation for predicted landmarks and a median error of \SI{2.72}{[\m\metre]}.
The benchmark methods UPF, INRIA, and CarMEN present median errors of \SI{3.17}{[\m\metre]}, \SI{4.88}{[\m\metre]}, and \SI{3.81}{[\m\metre]}, respectively. 
WarpPINN and WarpPINN-fibers achieve similar results to UPF, and outperform INRIA and CarMEN. In Figure~\ref{fig:boxplot_mesh_lmks} we also visualize the predicted deformed state of the left ventricle of volunteer v1 at end-systole, with the ground-truth landmarks (in green dots) and the predicted landmarks (in red dots) across all models. Here, it can be observed that the predictions from warpPINN and WarpPINN-fibers tend to produce uniformly thickened ventricular walls at end-systole, as opposed to the other methods, which exhibit decreased thickness in the apical region and some parts of the base and mid-section of the ventricles (in the case of UPF and CarMEN) or present irregularities throughout the entire domain (in the case of INRIA).

\begin{figure}[b]
    \centering
    \includegraphics[scale = 0.6]{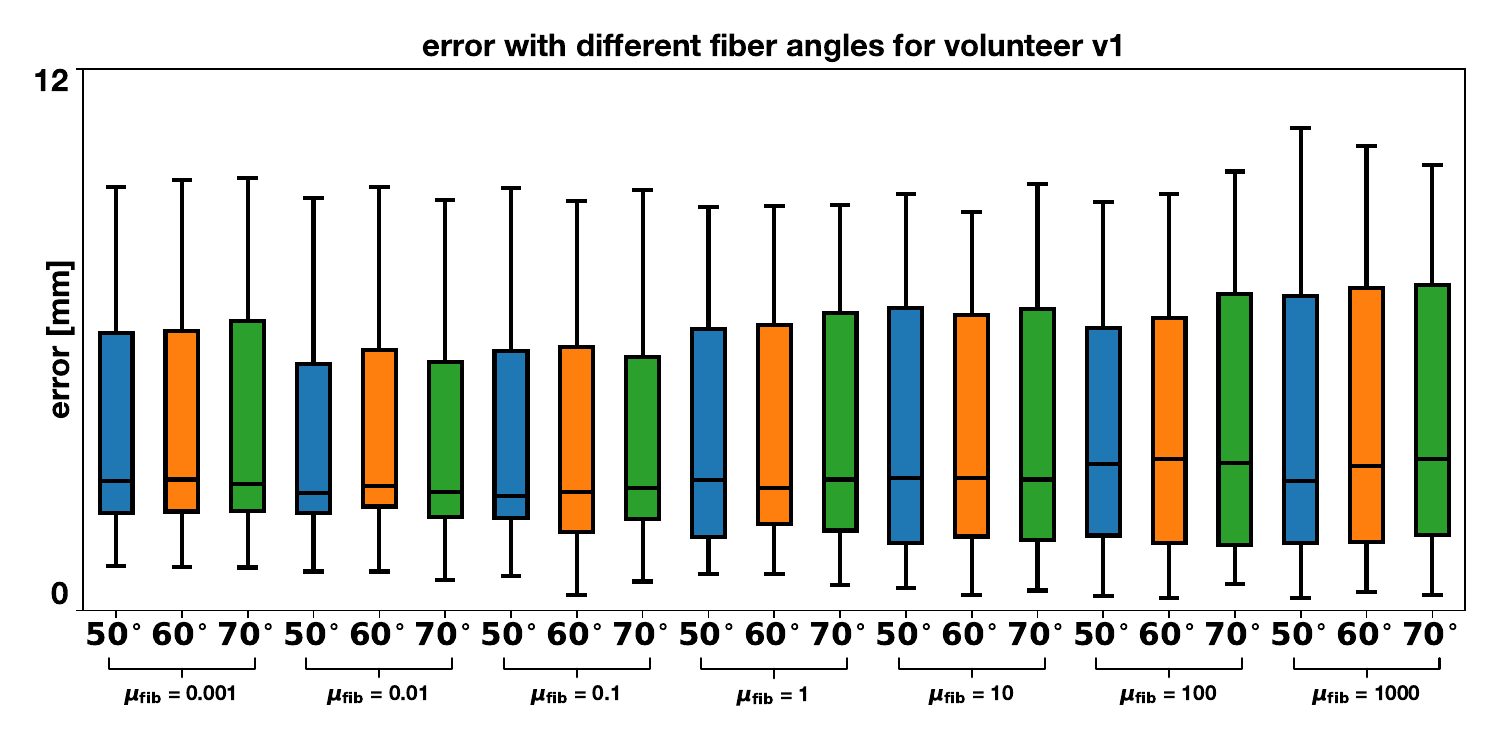}
    \caption{Box plots for landmark tracking errors on v1 using WarpPINN-fibers with different combinations of fiber angle $\alpha$ ($50^{\circ}, 60^{\circ}$ and $70^{\circ}$) and $\mu_{\mathbf{}}$.}
    \label{fig:error_mu_fib_alpha}
\end{figure}

Although the WarpPINN-fibers approach excels alternative methods in the task of landmark tracking, it becomes necessary to demonstrate its sensitivity to the fiber angle $\alpha$ selected and the weight $\mu_\mathbf{f}$ assigned to the fiber stretch loss. To evidence the proficiency of our method over different hyper-parameter configurations, we show in Figure~\ref{fig:error_mu_fib_alpha} a boxplot that measures the accuracy of our network for v1 over multiple combinations of $\alpha$ and $\mu_\mathbf{f}$. We observe that predictions for landmark location present small variations despite notable changes in the helix angle $\alpha$ within the range of histological studies \cite{lombaert2012}.

\begin{figure}[t]
	\centering
	\includegraphics[width=\textwidth]{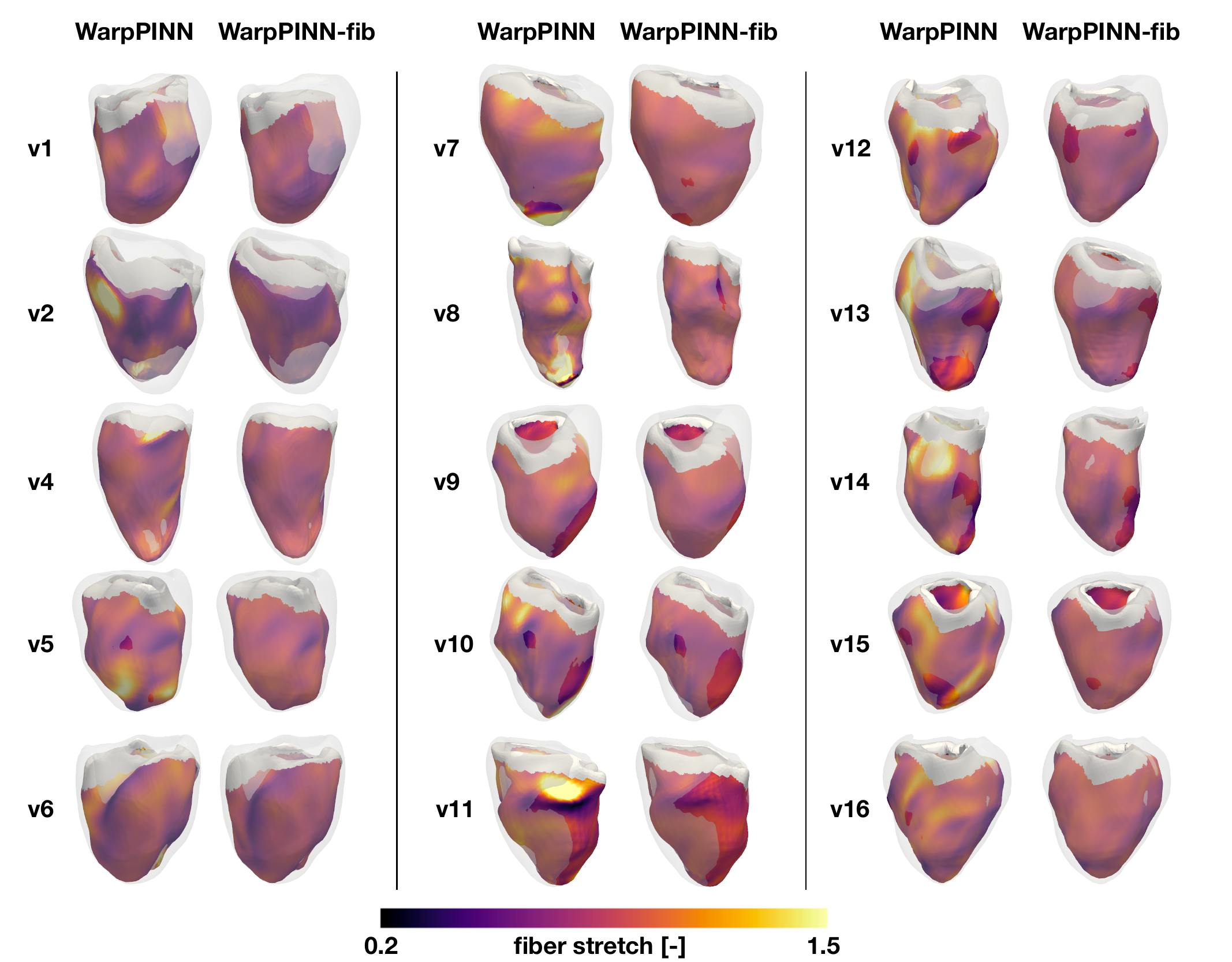}
	\caption{Deformed configurations at end-systole for WarpPINN and WarpPINN-fibers (WarpPINN-fib) with corresponding fiber stretch for each volunteer. The base of each ventricle is shown in gray. Each geometry is overlaid with its end-diastolic configuration in translucent gray.}
	\label{fig:LV_fib_stretch}
\end{figure}

For a detailed comparison between the physics-informed neural network approaches, in  
Figure~\ref{fig:LV_fib_stretch} we examine the differences between end-diastolic and end-systolic configurations and show the fiber stretch predicted at end-systole for all volunteers. WarpPINN-fibers tends to produce regions with less fiber stretch than WarpPINN, while inducing similar but more constrained deformations near the mid-cavity.
Interestingly, even though we do not penalize WarpPINN-fibers at the base, our method is still able to control deformation within that area, as observed in volunteers v1, v2, v6, v8, v12 and v13.

\begin{figure}[t]
	\centering
	\includegraphics[width=\textwidth]{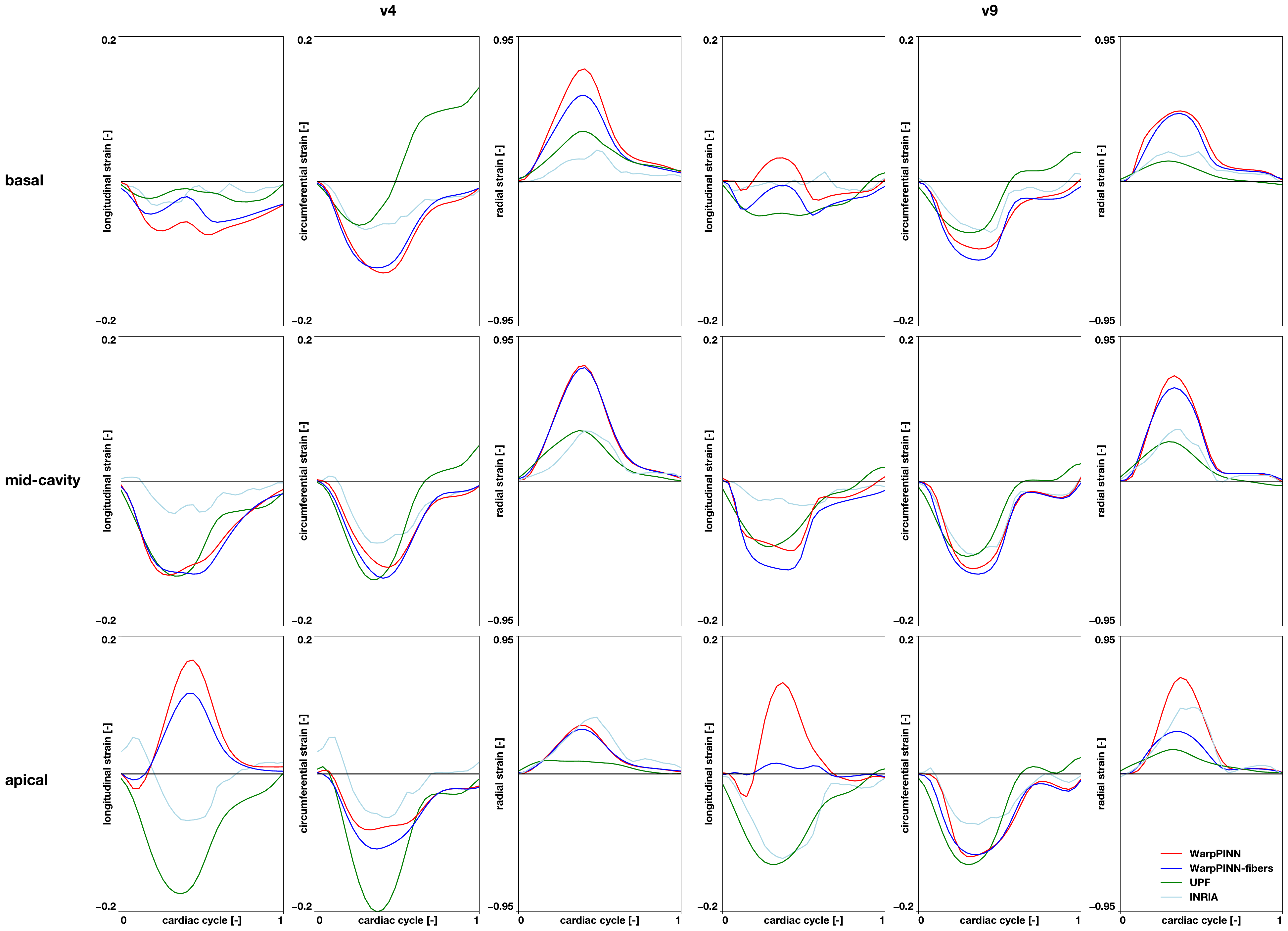}
	\caption{Strain curves predicted by WarpPINN (red; $\mu = 10^{-5}$, $\sigma = 1$), WarpPINN-fibers (blue; $\mu = 10^{-5}$, $\sigma = 1$, $\mu_{\mathbf{f}} = 10$), UPF (green) and INRIA (light blue) for volunteers v4 and v9. The first, second and third columns represent longitudinal, circumferential and radial strains, respectively. Results are divided in basal (top), mid-cavity (center) and apical (bottom) regions.}
	\label{fig:strain_comparison}
\end{figure}

Finally, Figure~\ref{fig:strain_comparison} depicts the average longitudinal, circumferential, and radial strains predicted by WarpPINN, WarpPINN-fibers, UPF, and INRIA over the cardiac cycle for volunteers v4 and v9.
These results are grouped according to the AHA segments, covering the basal, mid-cavity and apical sections of the left ventricle. We see that strain curves have similar behavior for most cases, but at different scales. These curves evolve smoothly over the cardiac cycle, with the exception of INRIA. In the basal region, the WarpPINN and WarpPINN-fibers present increased radial strain. Our fiber-based approach shows a distinct behavior of longitudinal strain in v9, having negative values for this component at the base. This suggests that WarpPINN-fibers predicts a mitral annular motion towards the apex, which is not the case for WarpPINN. In the mid-cavity region, most methods are in agreement for the quantification of longitudinal and circumferential strains, but have notable differences in the radial direction, where both WarpPINN and WarpPINN-fibers produce higher values compared to the other approaches. In the apical section, the strain curves from WarpPINN and WarpPINN-fibers differ in sign from both UPF and INRIA in the longitudinal direction. In this case, the strain predicted by WarpPINN-fibers remains below the WarpPINN approach, having a marked difference in patient v9. This behavior shows that, near the apex, WarpPINN-fibers enforces expansion in the vertical direction.

\section{Discussion}
\label{sec:discussion}

In this work, we introduced WarpPINN-fibers, a physics-informed neural network architecture designed for the reconstruction of cardiac strains in the image registration problem for cine-MR images. Based on the WarpPINN \cite{lopez2023warppinn} model, WarpPINN-fibers extends the notion of almost-incompressible cardiac tissue to an enhanced mechanical model that exploits the influence of myofibers into cardiac motion without requiring additional data. This is done by incorporating a fiber-stretch penalization over computationally generated fibers. 

From a synthetic example consisting of a cylindrical phantom model subjected to compression and torsion, we evidence that WarpPINN-fibers significantly improves the prediction of deformation fields while adjusting fiber shortening across the cardiac cycle. We demonstrate that the inclusion of a fiber penalization term favors realistic contraction by making the right Cauchy-Green deformation tensor projected onto the fiber direction at most $1$. This approach prevents the anomalous behavior produced by WarpPINN near the boundary of the phantom, specifically on the edges of the top and bottom caps of the cylinder. We argue that this issue may be attributed to the resolution used in the $z-$axis, as the images were composed of stacks of 21 slices in that direction. Although showing some discrepancy with the exact solution near the same regions, our fiber-informed model constrains the deformation estimate and notably improves error on both bases of the cylinder. Moreover, the substantial increase in performance derived from WarpPINN-fibers is quantitatively supported by the corresponding MSE, which shows a 44\% reduction compared to the other method. 

From our application to the STACOM-2011 cine SSFP MR image dataset of 15 volunteers, we demonstrate that WarpPINN-fibers can control stretch in the direction of cardiac fibers to achieve more realistic ventricular deformation and strain curves that are aligned with the physiological motion of the heart. Even though the stretch penalization term is not applied at the base of the ventricle, our fiber-informed model propagates the anisotropic behavior of myofiber mechanics to produce a contractile motion in the mitral annular region. Additionally, while the mean fiber stretch values are similar between WarpPINN and WarpPINN-fibers, their maximum values differ significantly. This highlights the proficiency of our method to stabilize cardiac motion on sections where the mesh is irregular or the object is undefined because of low-image resolution.

From the strain curves obtained in the basal segments of the left ventricle, we deduce that the WarpPINN-fibers method is a more reliable predictor of radial and longitudinal strains than alternative methods. Moreover, the negative values for longitudinal strain obtained with our approach are consistent with the observed displacement of the mitral annular plane towards the apex during systole \cite{MATOS2012969,wang2023,ballesterRodes2006}. In contrast, our method outputs positive strains in the longitudinal direction for the apical sections of the ventricle. However, this is expected due to the way the longitudinal direction is defined: since we set it to be a uniform vertical field $\mathbf{l}:=(0,0,1)^T$, the ventricular wall becomes progressively orthogonal to this direction as we move closer to the apex. This makes systolic wall thickening \cite{bart1972,prasad2010} vertically oriented, thus producing a change to positive sign as we approach the apical segments. Furthermore, this component is reduced in magnitude compared to the WarpPINN approach. Although the results from our method align better with the typically observed behavior of the apex during contraction, \cite{ballesterRodes2006, CODREANU201048}, they may be occasional and possibly improved by increasing the resolution in the long axis, which was of approximately \SI{8}{[\m\metre]} in our case. 

In the remainder of the ventricle, strain patterns across time were consistent with clinical reports \cite{moore2000three}, with negative values predicted for the longitudinal and circumferential strains and positive values for the radial strain. The comparatively higher radial strain obtained with the physics-informed architectures is explained by the increased wall thickening observed during systole with both WarpPINN and WarpPINN-fibers. This phenomenon is in agreement with reports from other studies \cite{bart1972, GERMANO19971360, prasad2010, ordas2005}, and manifests more prominently in the radial component, as it is computed from the normal field of the geometry.

An essential advantage of our WarpPINN-fibers pipeline is that it does not required additional measurements to construct the fiber loss, since fiber directions are synthetically produced with a Laplace-Dirichlet rule-based (LDRB) algorithm that only requires a user-specified helix angle $\alpha$. We demonstrate that our method is robust to different values of $\alpha$ within the range of histological observations \cite{lombaert2012} over several weights for penalization. Moreover, our method has the potential to incorporate more detailed models for artificial fiber generation and could be easily adapted to integrate patient-specific fiber distribution from more sophisticated imaging techniques, since our stretch penalization only depends on the predicted deformation and the fiber direction available.

Even though WarpPINN-fibers outperforms alternative methods in both synthetic and applied settings, successfully controlling myocytic stretching in favor of physiologically accurate contraction of tissue, our study is limited in the analysis of strain curves. Since ground-truth displacements are unavailable in the dataset used for our assessment, our capability to fully quantify the proficiency of our method to capture real cardiac motion becomes restricted. Measuring the accuracy for the magnitude of strain remains difficult and the literature presents mixed results. One study reports averages for longitudinal, radial and circumferential strain of $-0.1$, $0.4$ and $-0.1$, respectively, for DENSE \cite{cao2018comparison}, while other study reports magnitudes below $0.2$ for average values of strain components \cite{moulin2021myofiber}. Tracking methods present mean values in the range of $0.23$ to $0.64$ for radial strain and $-0.16$ to $-0.10$ for circumferential strain \cite{cao2018comparison}. Tagging techniques display mean values similar to feature tracking methods \cite{cao2018comparison,moore2000three}. Furthermore, strain measurements are not only influenced by the short-axis segmentation considered, but also the wall section selected for strain quantification (anterior, lateral, inferior or septal) \cite{moore2000three}.



In summary, the WarpPINN-fibers is a novel physics-informed methodology that incorporates fiber mechanics into a quasi-incompressibility model to accurately reconstruct cardiac motion from cine-MR images. Our method is robust and more accurate than other registration techniques. We believe WarpPINN-fibers will serve as a more reliable tool to assess cardiac function without demanding resources beyond standard MR scanners and reasonable time for reconstruction.



 


\section{Acknowledgments}
FAB, TB, IS and FSC acknowledge the financial support of the project ERAPERMED-134 from ANID. TB and FSC also acknowledge the support of the Millennium Institute for Intelligent Healthcare Engineering i-Health, ICN2021\_004.



\bibliographystyle{elsarticle-harv}
\bibliography{sn-bibliography}

\appendix

\section{Synthetic images}
\label{app:synth_images}

\begin{figure}[H]
	\centering
	\includegraphics[width=\textwidth]{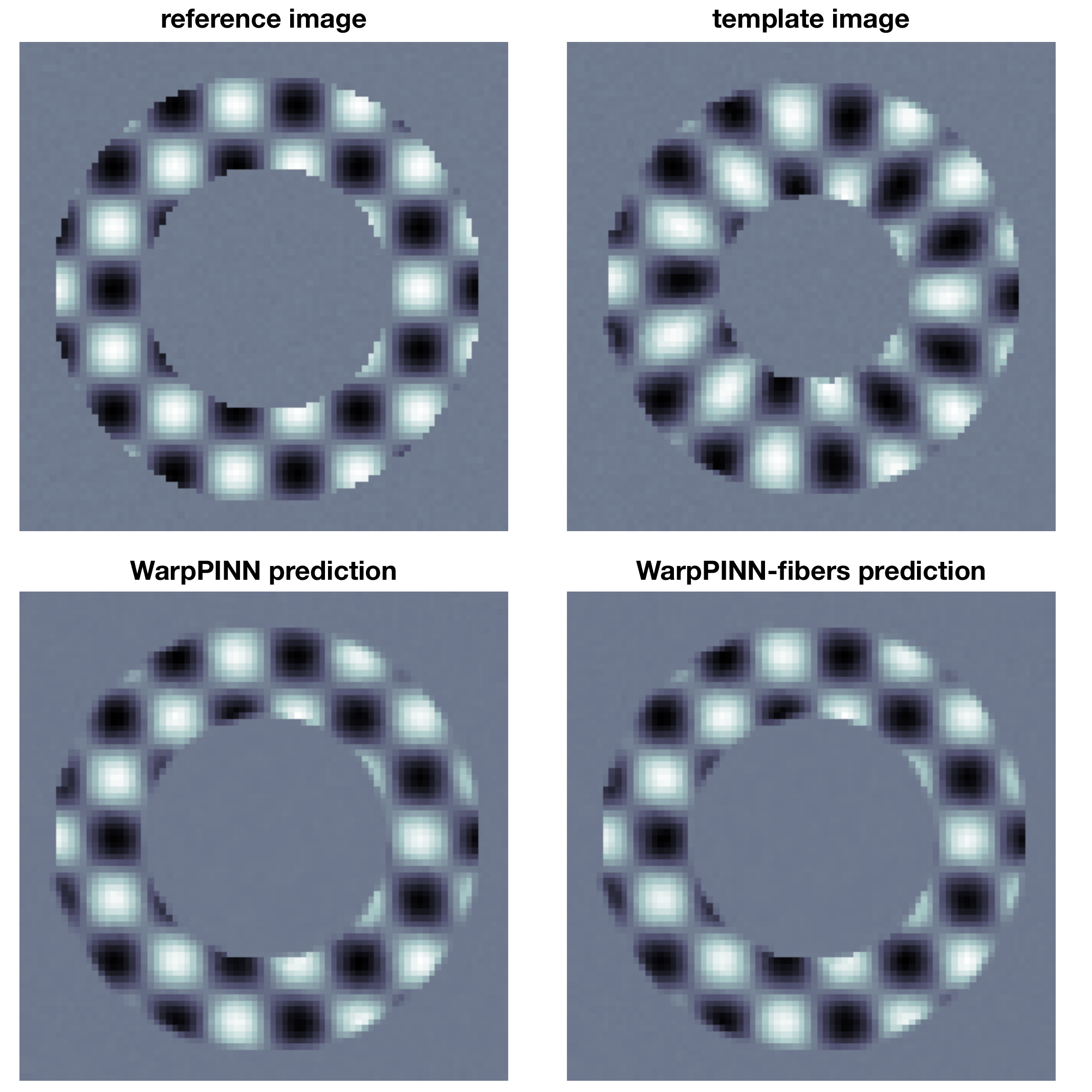}
	\caption{Image reconstruction for synthetic example in a horizontal cut at $z=0$. On the top row we show the reference (left) and template images (right) and on the bottom row we show the predicted deformation composed with the template image for WarpPINN (left) and WarpPINN-fibers (right).}
	\label{fig:synth_images}
\end{figure}

For the synthetic example, we generate a reference image of the cylinder at rest and a template image of the cylinder under compression and torsion. The top row of Figure \ref{fig:synth_images} shows a cut in the $xy$ plane at $z=0$ for both the reference and template images. In the bottom row, the predictions from WarpPINN and WarpPINN-fibers are displayed. They show good agreement with the ground truth.

\section{Fourier feature mapping hyper-parameters}
\label{app:FF_sigma_m}

\begin{figure}[H]
	\centering
	\includegraphics[width=0.45\textwidth]{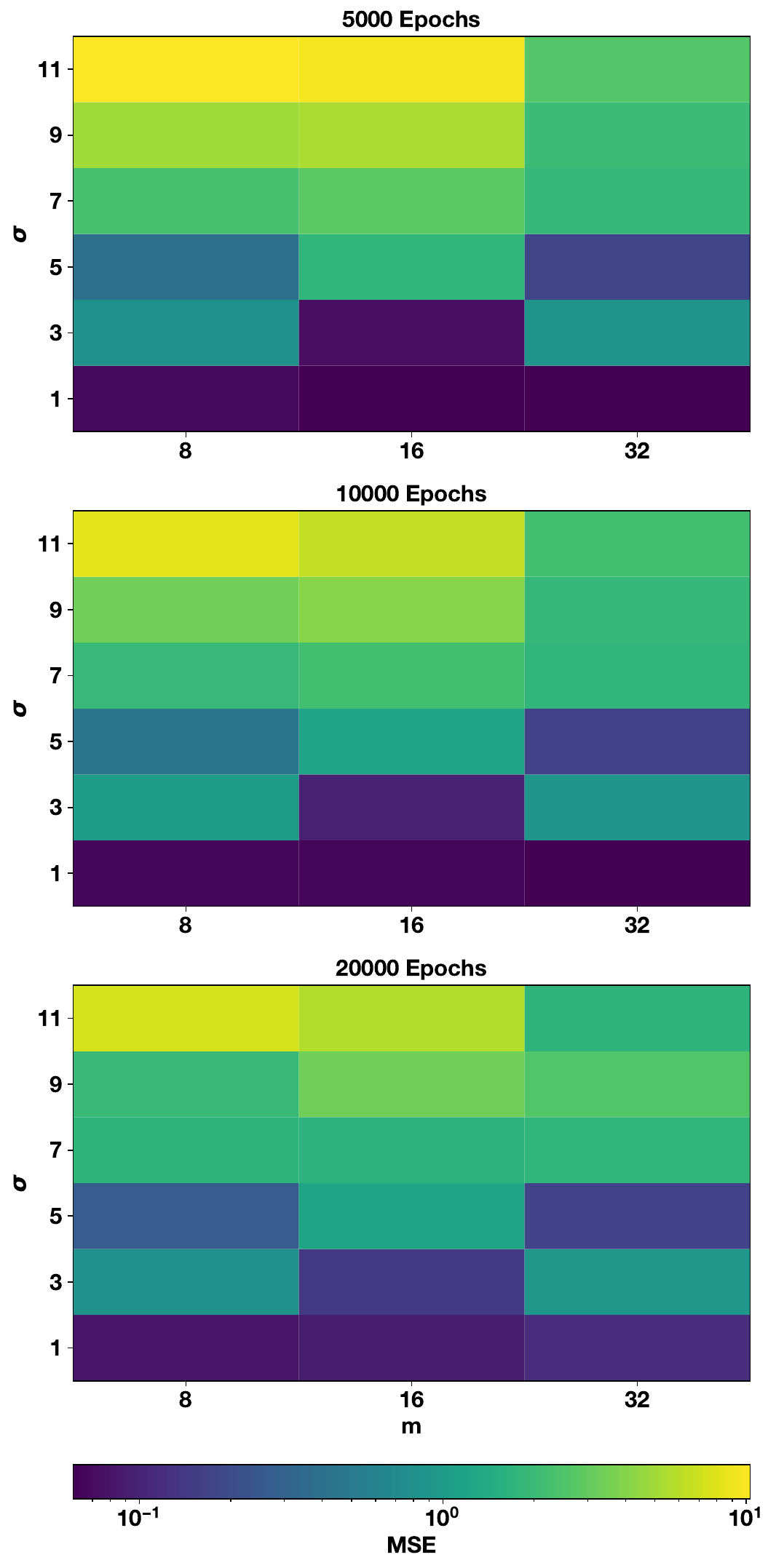}
	\caption{Mean squared error (MSE) of WarpPINN-fibers in the synthetic experiment for different combinations of the Fourier feature mapping hyper-parameters $m$ and $\sigma$. We present these results as heat maps corresponding to 5,000 (top), 10,000 (center) and 20,000 (bottom) training epochs.}
	\label{fig:synth_sigma_vs_m}
\end{figure}

We test WarpPINN-fibers on the synthetic example for 5,000, 10,000 and 20,000 epochs using various combinations of $m$ and $\sigma$ for the FF mappings.
The mean squared errors are computed using the known displacement field and are presented in the pseudo-color plots from Figure \ref{fig:synth_sigma_vs_m}.
For each set of simulations, the lowest MSE is obtained with $\sigma = 1$. Specifically, the minimal MSE values attained with our method are $6.05 \cdot 10^{-2}$ for 5,000 epochs ($m = 16$), $6.10 \cdot 10^{-2}$ for 10,000 epochs ($m = 32$) and $\text{MSE} = 8.02 \cdot 10^{-2}$ for 20,000 epochs ($m = 8$).

When higher values of $\sigma$ are considered, we observe that our method tends to predict displacement with increased error. However, there is no clear trend for MSE when $m$ increases. Additionally, for high values of $\sigma$, MSE decreases as more epochs are used for training.

\end{document}